%% file: Uncertainty_Decomposition.tex
\documentclass[conference]{IEEEtran}
\IEEEoverridecommandlockouts
\usepackage{cite}
\usepackage{amsmath,amssymb,amsfonts}
\usepackage{algorithmic}
\usepackage{graphicx}
\usepackage{textcomp}
\usepackage{xcolor}
\usepackage{booktabs}
\usepackage{amsmath}
\usepackage{subcaption}
\usepackage{makecell}
\begin{document}

\title{Uncertainty Decomposition and Error Margin Detection of Homodyned-K Distribution in Quantitative Ultrasound\\
\thanks{Funded by the Government of Canada’s New Frontiers in Research Fund
(NFRF), [NFRFE-2022-00295] and Natural Sciences and Engineering Research Council of Canada (NSERC).}
}

\author{\IEEEauthorblockN{ Dorsa Ameri}
\IEEEauthorblockA{
\textit{Concordia University}\\
Montreal, Canada\\
Dorsa.Ameri@mail.concordia.ca}
\and
\IEEEauthorblockN{Ali K. Z. Tehrani}
\IEEEauthorblockA{
\textit{Concordia University}\\
Montreal, Canada\\
ali.kafaeizadtehrani@mail.concordia.ca}
\and
\IEEEauthorblockN{Ivan M. Rosado-Mendez}
\IEEEauthorblockA{
\textit{University of Wisconsin}\\
Madison, United States\\
rosadomendez@wisc.edu}
\and
\IEEEauthorblockN{Hassan Rivaz}
\IEEEauthorblockA{
\textit{Concordia University}\\
Montreal, Canada\\
hrivaz@ece.concordia.ca}
}
\maketitle
\begin{abstract}
Homodyned K-distribution (HK-distribution) parameter estimation in quantitative ultrasound (QUS) has been recently addressed using Bayesian Neural Networks (BNNs). BNNs have been shown to significantly reduce computational time in speckle statistics-based QUS without compromising accuracy and precision. Additionally, they provide estimates of feature uncertainty, which can guide the clinician's trust in the reported feature value. The total predictive uncertainty in Bayesian modeling can be decomposed into epistemic (uncertainty over the model parameters) and aleatoric (uncertainty inherent in the data) components. By decomposing the predictive uncertainty, we can gain insights into the factors contributing to the total uncertainty. In this study, we propose a method to compute epistemic and aleatoric uncertainties for HK-distribution parameters ($\alpha$ and $k$) estimated by a BNN, in both simulation and experimental data. In addition, we investigate the relationship between the prediction error and both uncertainties, shedding light on the interplay between these uncertainties and HK parameters errors. 
\end{abstract}

\begin{IEEEkeywords}
Homodyned K-distribution, Quantitative Ultrasound, Bayesian Neural Networks,  Uncertainty Estimation
\end{IEEEkeywords}
\section{Introduction}
Scatterers are microstructures within the tissue that are smaller than the ultrasound wavelength and scatter the ultrasound waves. Quantitative ultrasound (QUS) analyzes the detected backscattered signal to provide insights into the scatterers' structures, which are closely linked to the tissue characteristics. QUS methods have been widely utilized for tissue characterization, including liver fibrosis staging \cite{zhou2020value,bhatt2021multiparametric, timana2023simultaneous}, breast inclusions classification \cite{muhtadi2022breast,chowdhury2022ultrasound,destrempes2020added,khairalseed2023high}, and metastatic lymph nodes detection \cite{hoerig2023classification}.

QUS methods can be decomposed into two broad categories: spectral-based and envelope-based methods \cite{kim2008hybrid}. Spectral-based methods analyze the backscattered signal in the frequency domain to obtain parameters such as backscattering coefficient, attenuation coefficient, and effective scatterer diameter \cite{oelze2016review,jafarpisheh2023physics}. 
Envelope-based methods utilize the envelope of the backscattered signal to estimate QUS parameters, such as scatterer number density, and coherency of scatterers \cite{cloutier2021quantitative,oelze2016review,rivaz20079c}. These methods model the envelope of the backscattered RF data by fitting a distribution to the samples \cite{oelze2016review}. 

The Homodyned-K distribution can model the envelop data, and its parameters have been found to be correlated with the tissue properties, making them useful in tissue characterization \cite{bhatt2021multiparametric,byra2016classification}. 
Envelope statistics, including point-wise signal-to-noise ratio (SNR), skewness, kurtosis, and log-based moments, are commonly used to estimate HK parameters. Estimating HK parameters from known envelope statistics does not have a closed-form solution, and conventional methods of estimating these parameters rely on iterative optimization methods or table search \cite{destrempes2013estimation, Hruska2009, liu2023study}.

Deep learning methods have recently been employed in QUS \cite{oh2022sensor, tehrani2023deep}. They have also been utilized to estimate the parameters of the HK distribution by employing envelope statistics as input features of the model.
Zhou et al. introduced an Artificial Neural Network (ANN) as an estimator of HK parameters \cite{zhou2021parameter}. The ANN estimator employed Multilayer Perceptrons (MLPs), which were prone to overfitting. Moreover, ANNs perform as a black box without having a reliability metric, which is crucial to ensure that the model's predictions are trustworthy and can be confidently applied in clinical or research settings.
Tehrani et al. addressed these issues by utilizing a Bayesian Neural Network (BNN) which incorporates probability distributions instead of fixed weights, enabling the estimation of the uncertainty of predictions. Their method improved the estimation of HK-distribution parameters and provided uncertainty quantification which enables the assessment of the reliability of the model's outputs \cite{tehrani2024homodyned, tehrani2022homodyned}.

The total predictive uncertainty in Bayesian modeling can be decomposed into epistemic and aleatoric components\cite{kendall2017uncertainties}. 
Epistemic uncertainty refers to the uncertainty over the network weights. This uncertainty is often referred to as \textit{model uncertainty}.  High epistemic uncertainty indicates that the test input may be an outlier or different compared to the training distribution. As a result, by training the model with more diverse training data, we can reduce this type of uncertainty.
Aleatoric uncertainty, on the other hand, captures the noise inherent in the input observations. Addressing this type of uncertainty requires knowledge about unobserved variables, such as additional features in the input, which are often inaccessible. Therefore, it is not always possible to reduce aleatoric uncertainty \cite{kendall2017uncertainties,chai2018uncertainty}.
By understanding the contributions of the two uncertainty components, we can identify whether increasing the size of the training data or improving the data collection process would be beneficial to reduce the total uncertainty.



In this study, we propose a framework to obtain the total predictive uncertainty of HK-distribution parameters and decompose it into epistemic and aleatoric ones using BNN for simulation and experimental data. We also investigate the relationship between the prediction error and both uncertainties.

\section{Materials and Methods}

\subsection{Homodyned K-distribution}
The Homodyned K-distribution (HK-distribution) is described by the following equation \cite{destrempes2013estimation}:

\begin{equation}
    \scalebox{0.95}{
        $P_{HK}(A|\epsilon ,\sigma^2,\alpha) = A\int_{0}^{\infty}uJ_{0}(u\epsilon)J_{0}(uA)\left(1+\frac{u^2\sigma^2}{2}\right)^{-\alpha}du$
    }
    \label{Eq1}
\end{equation}

where \( A \)  represents the envelope of the backscattered echo signal and \( J_0 (\cdot) \) is the zero-order Bessel function. The coherent signal power is denoted by \( \epsilon^2 \), and the diffuse signal power is given by \( 2\sigma^2\alpha \) \cite{destrempes2013estimation}. The scatterer clustering parameter $\alpha$, and coherent to diffuse scattering ratio $k = \frac{\epsilon}{\sigma \sqrt{\alpha}}$, referred here to as HK parameters, are commonly used in tissue characterization \cite{gesnik2020vivo,ghoshal2012ex}. $\alpha$ and \( k \) are correlated with the scatterer number density and the microstructural organization of the scatterers, respectively. 
\subsection{Datasets and data generation}

\paragraph{Simulation Data}
Equation \eqref{Eq2} suggested by \cite{hruska2009improved},\cite{zhou2021parameter} was employed to produce synthetic samples from HK-distribution:
\begin{equation}
\scalebox{0.95}{
    $a_i = \sqrt{\left( \sqrt{2k} + X_i \sigma \sqrt{Z_i/\alpha} \right)^2 + \left( Y_i \sigma \sqrt{Z_i/\alpha} \right)^2}$
    }
    \label{Eq2}
\end{equation}

\noindent where \(a_i\) is the generated sample, and \( X_i\) and \( Y_i\) are the independent and identically distributed samples (i.i.d) from the Normal distribution having zero mean and variance of 1. \( Z_i\) is the sample from the Gamma distribution with shape parameter $\alpha$ and scale parameter $\sigma$ which is set to 1.

We first trained the BNN following the schedule suggested by \cite{tehrani2024homodyned}. The simulation test data were generated for 31 randomly selected \(\log_{10}(\alpha)\) values ranging from -0.3 to 1.3 corresponding to $\alpha$ from 0.5 to 20, and 11 values of \( k \) ranging from 0 to 1.25 similar to \cite{tehrani2024homodyned} using \eqref{Eq2}. Moreover, for each value of \(\log_{10}(\alpha)\) and \( k \), 10 realizations were generated, leading to 3410 sample sets. Different input observations were used for uncertainty decomposition, which will be elaborated on in Subsection C.
In addition, Rayleigh noise was incorporated into the samples at different levels, resulting in data with three different SNRs of 20, 30, and 40 \text{dB}, which are computed by $ SNR=10 \log_{10} \left( E[\text{env}^2] / (2 \sigma_N^2) \right)$ where \(E[\text{env}^2]\) denotes the power of the envelope samples and \(\sigma_N\) is the scale parameter of Rayleigh noise.

\paragraph{Experimental Data} 
We used data from four experimental phantoms. The first dataset was sourced from a phantom previously described in \cite{nam2012comparison}, which is a three-layered phantom with two different scatterer number densities. The phantom was constructed using an emulsion of ultrafiltered milk and water-based gelatin. Glass beads with diameters ranging from 5 to 43 µm (3000E, Potters Industries, Valley Forge, PA, USA) were used as scattering sources. Images of the phantom were acquired using an 18L6 linear array transducer with a center frequency of 8.9 MHz on a Siemens Acuson S2000 scanner (Siemens Medical Solutions USA, Inc.), previously reported in \cite{nam2012comparison}.
The top and bottom layers of the phantom have the same scatterer concentration of 2 g/L, and the middle layer has a higher concentration of 8 g/L. Patches R1 and R2 were selected to extract the statistical features from the top and middle layers, respectively. We obtained 60 patches of envelope data for each of the two specified regions by moving the 
patches laterally (to avoid any changes in resolution cell size) across several frames. 

The other three datasets were acquired from homogeneous phantoms with different numbers of scatterers per resolution cell, previously reported in \cite{tehrani2021ultrasound}. 
The dimensions of the phantoms measured 15 cm × 5 cm × 15 cm, and they were made from a homogeneous mixture of agarose gel and glass beads as scattering agents. The diameter range of the glass beads and their concentration in the phantoms are detailed in \cite{tehrani2021ultrasound}. For further construction details, including the speed of sound and attenuation coefficient, see \cite{rosado2014advanced}. Data was collected with an 18L6 transducer operating at a 10 MHz frequency using an Acuson S2000 scanner (Siemens Medical Solutions, Malvern, PA, USA). The phantoms are referred to as Phantom A (high concentration), Phantom B (medium concentration), and Phantom C (low concentration) having scatterer concentrations of 236, 9, and 3 per $\textit{mm}^3$, respectively. Seventy-two patches were acquired from each phantom.

\subsection{Uncertainty Decomposition}
\begin{figure}[b]
\centerline{\includegraphics[width=0.23\textwidth]{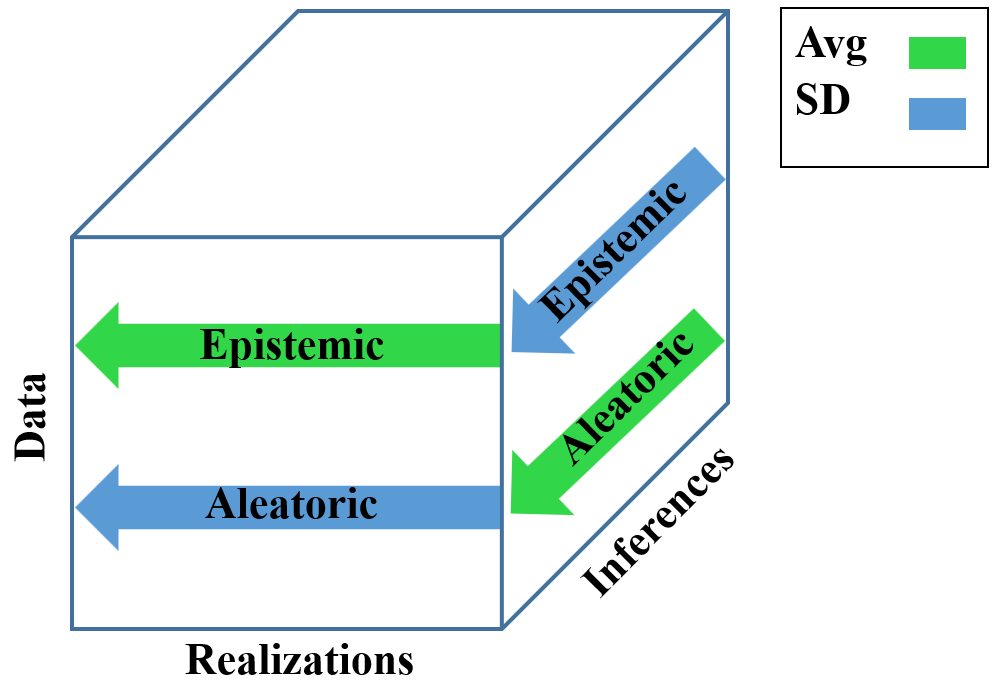}}
\caption{The visualization of computing the uncertainties.}
\label{fig1}
\end{figure}

To estimate the two components of the uncertainty in our model's predictions, we used the method proposed in \cite{lee2020calibrated}:

\begin{equation}
\scalebox{0.95}{
    ${Var}(y) \approx 
    \underbrace{E(\hat{y}^2) - E(\hat{y})^2}_\text{Epistemic Uncertainty} + 
    \underbrace{E(\hat{\sigma}^2)}_\text{Aleatoric Uncertainty}$
    }
    \label{Eq3}
\end{equation}

\noindent where $E(\hat{y})$ denotes the expected value of the prediction and $\hat{\sigma}^2$ is the predicted variance.  Expected values are obtained by Monte Carlo sampling at inference time.
According to \eqref{Eq3}, to estimate each uncertainty component for the simulation data, different observations of the input data and the model weights were required. The trained BNN was used in inference with 50 times sampling of weights for each of the 10 input observations for simulation data, yielding $50\times10$ different output estimates.
Having the total test data \( dt \in \mathbb{R}^{341 \times 10} \), the dimensions of the total prediction were \( k, \log_{10}(\alpha) \in \mathbb{R}^{341 \times 10 \times 50} \), where the last two dimensions correspond to different input realizations and different BNN inferences, respectively.
As shown in \eqref{Eq3}, the total predictive uncertainty is the summation of epistemic and aleatoric uncertainties. The epistemic uncertainty was estimated by computing the standard deviation over the 50 model inferences,  followed by averaging these deviations over the predictions of 10 different input realizations. On the other hand, the aleatoric uncertainty was estimated by first averaging the predictions across the 50 model inferences, and then computing the standard deviation over the predictions of 10 input realizations. This captures the variations across different realizations and removes the uncertainty of the model in different inferences.
The visualization of the calculations is illustrated in Fig.~\ref{fig1}.
A similar approach was followed to compute the epistemic and aleatoric uncertainties of the estimated values of $log_{10}(\alpha)$ in the experimental data, using different input observations (60 for layered phantom and 72 for homogeneous phantoms) and 50 model inferences.

\section{Results}
\begin{figure} [t]
  \centering
  \begin{minipage}{0.21\textwidth}
    \centering
    \includegraphics[width=\textwidth]{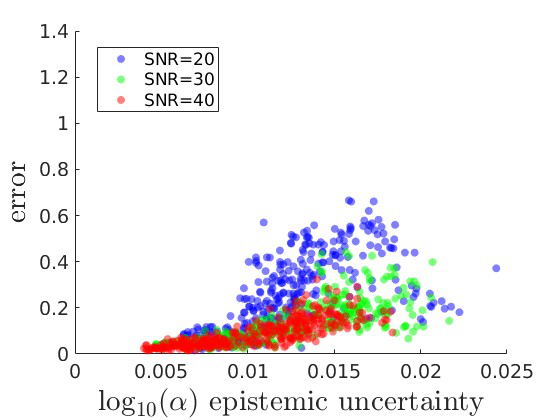}
    \subcaption{}
    \label{fig:subfig1_i}
  \end{minipage}%
  \begin{minipage}{0.21\textwidth}
    \centering
    \includegraphics[width=\textwidth]{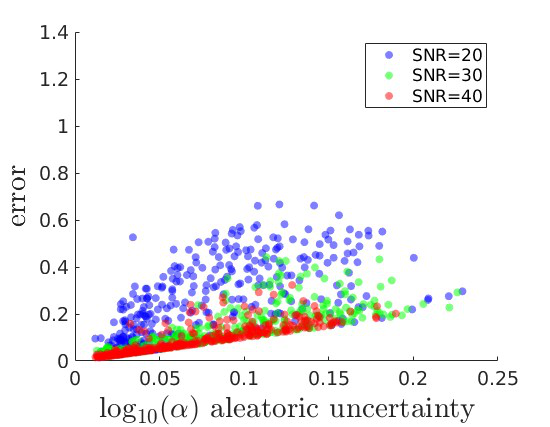}
    \subcaption{}
    \label{fig:subfig2_i}
  \end{minipage}%
  
  \begin{minipage}{0.21\textwidth}
    \centering
    \includegraphics[width=\textwidth]{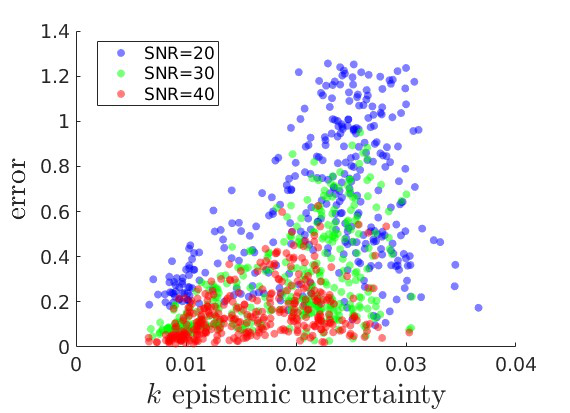}
    \subcaption{}
    \label{fig:subfig3_i}
  \end{minipage}%
  \begin{minipage}{0.21\textwidth}
    \centering
    \includegraphics[width=\textwidth]{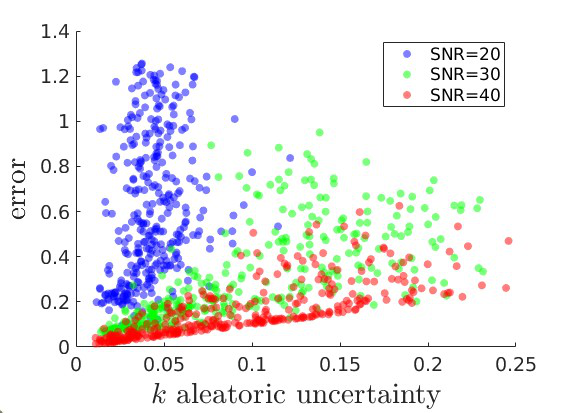}
    \subcaption{}
    \label{fig:subfig4_i}
  \end{minipage}
  \caption{Error versus epistemic and aleatoric uncertainty for simulation data
    \textbf{(a)}~Epistemic uncertainty for  \(\log_{10}(\alpha)\).
    \textbf{(b)}~Aleatoric uncertainty for  \(\log_{10}(\alpha)\).
    \textbf{(c)}~Epistemic uncertainty for \( k \).
    \textbf{(d)}~Aleatoric uncertainty for \( k\).
    }
  \label{fig2}
\end{figure}

The Root Mean Square Error ($RMSE = \sqrt{\frac{1}{n} \sum_{i=1}^{n} (\hat{y}_i - y_i)^2}$) of the model's predictions is calculated where \( n \) represents the number of data points, \(y_i\) and $\hat{y}_i$ denote the ground truth labels and model's predictions, respectively.
Fig.~\ref{fig2} illustrates the scatter plots representing the error values versus each uncertainty (epistemic and aleatoric) for estimated parameters \(\log_{10}(\alpha)\) and \( k \) for the simulation data with different SNRs.  
The correlation coefficients between the error and each uncertainty component for the simulation data at various SNR levels are provided in Table~\ref{tab1}. All \( p \)-values were below 0.01, indicating the statistical significance of the reported correlation values.
According to the plots in Fig.~\ref{fig2} and the correlation values in Table~\ref{tab1}, a relatively high correlation between the error and each uncertainty was observed. For data with SNRs of 30 and 40, aleatoric uncertainty shows a stronger correlation with error than epistemic uncertainty for both parameters \(\log_{10}(\alpha)\) and \( k \). On the other hand, for data with an SNR of 20, epistemic uncertainty demonstrates a higher correlation with error compared to aleatoric one, again for both parameters. Based on Fig.~\ref{fig2} (b) and (d), there is an apparent lower bound for the error as a function of the aleatoric uncertainty. Determining the lower bound for the error can be crucially important in tissue characterization since it allows researchers to acquire the minimum prediction error values without knowing the ground truth.

\begin{table}[t]
    \centering
    \caption{Simulation data Pearson’s correlation between error and uncertainties of predicted parameters (SNR = 20, 30, 40) }
    \resizebox{.4\textwidth}{!}{
    \begin{tabular}{lccc}
        \hline
        \textbf{Parameters} & \textbf{Epistemic} & \textbf{Aleatoric} & \textbf{Total} \\
        \hline
        $\log_{10}(\alpha)$ & 0.66, 0.75, 0.78 & 0.63, 0.80, 0.84 & 0.64, 0.81, 0.85 \\
        $k$ & 0.44, 0.52, 0.38 & 0.24, 0.71, 0.75 & 0.36, 0.72, 0.75 \\
        \hline
    \end{tabular}}
    \label{tab1}
\end{table}

Table~\ref{tab2} shows the uncertainties for different values of estimated \(\log_{10}(\alpha)\) in the experimental phantoms. The analysis of uncertainty values in the plots in Fig.~\ref{fig2} and Table~\ref{tab2} (for simulation and experimental data, respectively), indicates that aleatoric uncertainty constitutes a larger proportion of the total uncertainty than the epistemic one.
Therefore, adding more training data would not significantly reduce the overall predictive uncertainty here. If data acquisition is feasible, strategies such as using larger patches and angular compounding may help reduce aleatoric uncertainty, resulting in total uncertainty reduction. 

\begin{table}[b]
    \centering
    \caption{Aleatoric and epistemic uncertainties for predicted $log_{10}(\alpha)$ of the investigated experimental phantoms.}
    \resizebox{.42\textwidth}{!}{
    \begin{tabular}{lcccc}
        \hline
        \textbf{Phantoms} & \textbf{Epistemic} & \textbf{Aleatoric} &
        \makecell{\textbf{Interquartile range [$25\%, 75\%$]} \\ \textbf{of predicted $log_{10}(\alpha)$}} \\
        \hline
        Layered-R1 & 0.012 & 0.125 & [0.854, 1.034]\\
        Layered-R2 & 0.013 & 0.063 & [1.081,1.126]\\ \hline
        A (High) & 0.013 & 0.210 &[0.723, 1.074]\\
        B (Medium) & 0.006 & 0.137 & [0.387, 0.574]\\
        C (Low) & 0.010 & 0.173 & [-0.311, -0.137]\\
    \end{tabular}}
    \label{tab2}
\end{table}


\section{Discussion and Conclusion}
Previous studies that quantified the uncertainty of HK parameter estimations [24, 25] mainly focused on epistemic uncertainty, as they only accounted for uncertainty over the model parameters. 
Moreover, Tehrani et al. in \cite{tehrani2023deep} estimated an uncertainty that is similar to aleatoric uncertainty. In this study, we acquired both uncertainties and investigated their correlation with error.

Calculating aleatoric uncertainty is  challenging, as it requires multiple observations of the input data, which may not always be available. In contrast, epistemic uncertainty is easier to obtain through multiple inferences of the model in Bayesian frameworks. Our results in Section III show that the correlation between error and epistemic uncertainty is comparable to that between error and aleatoric uncertainty. This is valuable since epistemic uncertainty is always accessible, whereas aleatoric uncertainty may not be, especially in \textit{in vivo} data where obtaining multiple observations from the same tissue is difficult.

In this study, we introduced a framework to quantify and decompose the predictive uncertainty into epistemic and aleatoric components for Homodyned K-distribution (HK-distribution) parameters using Bayesian Neural Networks (BNNs).
Our results showed that the main contributor to the total uncertainty of the BNN model's predictions is the aleatoric uncertainty. Moreover, investigating the relationship between prediction errors and each uncertainty component leads to identifying a lower bound for the error values.




\input{uncertainty_decomposition.bbl}

\end{document}